# Reversible magneto-ionics in crystallized W-$Co_{20}Fe_{60}B_{20}$-MgO-$HfO_2$ ultra-thin films with perpendicular magnetic anisotropy

*Song Chen[1], Elmer Monteblanco[1], Benjamin Borie[1], Shimpei Ono[3], Dafiné Ravelosona[1,2]*

*[1]Spin-Ion Technologies, 10 Boulevard Thomas Gobert, 91120 Palaiseau, France*

*[2]Centre de Nanosciences et de Nanotechnologies, CNRS, Université Paris-Saclay, 10 Boulevard Thomas Gobert, 91120 Palaiseau, France*

*[3]International Center for Synchrotron Radiation Innovation Smart, Tohoku University, Aoba-Ku, Sendai 980-8572, Japan*

**Abstract**

We have investigated electric field (E-field) induced modulation of perpendicular magnetic anisotropy (PMA) in both amorphous and crystalline W/CoFeB/MgO/$HfO_2$ ultra-thin films. We find that in the amorphous state, the E-field effect is volatile and reversible, which is consistent with the conventional electrostatic effect through charge accumulation and depletion. In the crystallized system annealed at 370°C, we find that two effects are at play, a non-volatile and reversible voltage-induced effect on PMA and an electrostatic response. We discuss these results in terms of higher oxygen mobility at the crystallized CoFeB-MgO interface, which induces a non-volatile magneto-ionic response. Modulating PMA in crystallized CoFeB-MgO materials through ionic migration opens the path to integrating magneto-ionics in full magnetic tunnel junctions.

**Introduction**

Voltage control of magnetic properties such as interface magnetocrystalline anisotropy has been widely studied in ferromagnetic metal (FM)/metal oxide (MO) heterostructures due to its potential to reduce power consumption in spintronic devices.[1–4] The underlying physical phenomena are based on several magneto-electric mechanisms, including carrier-meditated electrostatic effect[5–10] or magneto-ionics response.[11–25] The conventional electrostatic effect induces a change in the relative occupancy of the orbitals of ferromagnetic metal atoms at the FM/MO interface, modifying the interface magneto-crystalline anisotropy and interface magnetization. However, the change in magnetic properties is limited due to electrostatic screening, and in addition, the effect is volatile. Magneto-ionics, where the voltage can modify the oxidation state of the FM material, has drawn intense interest recently since it provides non-volatility and a much larger change of magnetic properties with respect to pure charge effects. In particular, the influence of the FM/MO interface state on the MI effect has been widely studied, showing complex behavior in terms of modulation of magnetic properties (PMA, DMI…), reversibility, and non-volatility.[12,18,19,26,27]

However, poor attention has been devoted to the influence of the crystalline structure on the magneto-ionic response. Recently, it was shown that annealing-induced partial crystallization of W-CoFeB-$HfO_2$[12] and Ta-CoFeB-$HfO_2$[28] materials can result in a strong modification of the magneto-ionic response. In particular, it has been demonstrated that partial crystallization in W-CoFeB-$HfO_2$ can lead to a non-reversible MI response. However, full crystallization of CoFeB layers is a prerequisite in CoFeB-MgO-based Magnetic Tunnel Junctions (MTJ) to get a high TMR ratio for MRAM applications. It has also been recently observed that gating can also influence the crystalline structure of the MgO layer .[29]

Here, we demonstrate a reversible and non-volatile magneto-ionic effect in fully crystallized W-CoFeB-MgO materials with perpendicular anisotropy, which are the archetypical free layer in MTJs materials due to their large TMR, high anisotropy, low magnetic damping, and large Spin Hall effect. We use a 1 nm thick top $HfO_2$ dielectric layer that acts as an O reservoir to study voltage control of magnetic anisotropy (VCMA) and we compare the effect for both amorphous and crystallized W-CoFeB-MgO-$HfO_2$ materials. We show a volatile charge accumulation/depletion effect in the amorphous system, whereas, in the crystallized materials annealed at 370°C, the gate voltage can induce a reversible and non-volatile effect.

**Experimental Section**

The multilayer stack used in this study is based on W (4nm)/ $Co_{20}Fe_{60}B_{20}$ (1nm)/MgO (2nm)/ $HfO_2$ (1nm), structures grown by a Singulus Rotaris magnetron sputtering system on Si/$SiO_2$ substrates. The post-deposition annealing was implemented in a vacuum chamber at $1.0 \times 10^{-5}$ mbar at an annealing temperature of 370°C for one hour. To apply voltages, an area of 5 mm × 10 mm on top of the structure was covered by $[EMI]^+[TFSI]^-$ (1-ethyl-3-methyl imidazolium bis(trifluoromethanesulfonyl) imide), an ionic liquid (IL). A glass substrate coated with a 100 nm thick indium tin oxide (ITO) layer was used as the top electrode, while the bottom W electrode was connected to the ground as seen in Figure 1. ITO was connected to the contact pads on a PCB board. Hysteresis loops under a voltage were measured by polar magneto-optical Kerr effect using a Zeiss microscope at room temperature.

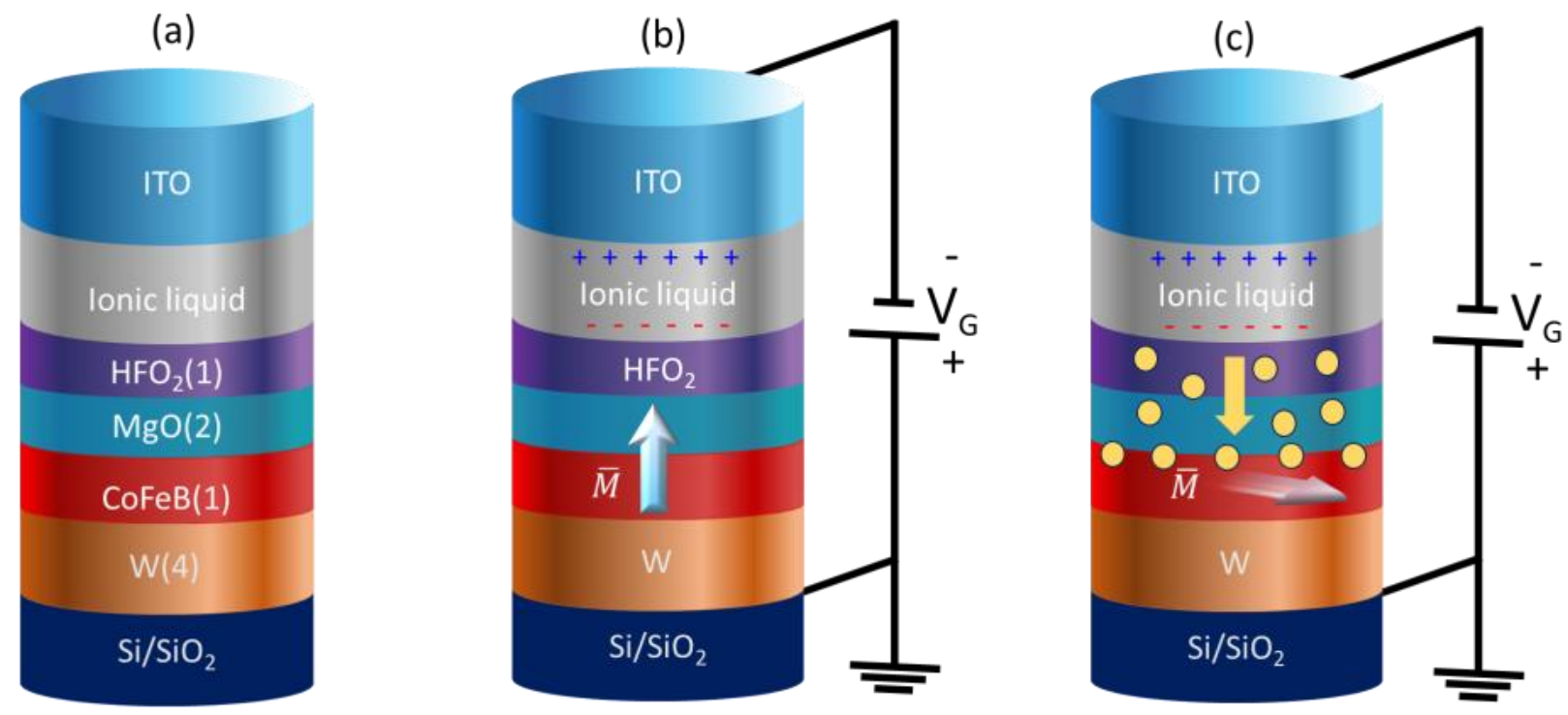

**Figure 1.** (a) Schematic structure of the $W/Co_{20}Fe_{60}B_{20}/MgO/HfO_2$ with ionic liquid and ITO glass as gate. (b)-(c) Gating voltage ($V_G$) applied between the ITO and the tungsten layer where + and – represent $[EMI]^+$ and $[TFSI]^-$. The flux of oxygen ions is represented by yellow circles. The transition from PMA to in-plane magnetized CoFeB is represented by the white arrow.

## Results and Discussion

### *Charge-meditated ME effect in as-deposited $W/Co_{20}Fe_{60}B_{20}/MgO/HfO_2$ materials*

Figure 2 (a) shows polar MOKE hysteresis loops under positive and negative voltages for non-annealed samples. The hysteresis loops were measured during the application of the voltage with a pulse duration of 180 s. The potential was maintained below +3.0V (-2.5V) to avoid inducing a high leakage current in the ionic liquid. The initial state, represented by the black dashed line in Figure 2 (a), corresponds to a magnetic state with low perpendicular magnetic anisotropy (PMA) where typical stripe domains are present. As depicted in Figure 2 (a), applying a negative voltage $V_G$ leads to an increase in the coercive field, while a positive E-filed $V_G$ results in a reduction. Notably, at the highest negative voltage, a perfect squareness of the hysteresis loop is achieved, corresponding to domain nucleation and propagation, as evidenced by the pMOKE images in Figures 2(b)-(c). The dendritic domain wall propagation suggests the presence of defects in the material. When a negative voltage (-2.5V) is applied to the system, domain nucleation initiates under a magnetic field of 1.16 mT. Conversely, under a continuously applied positive voltage of +3.0V, domain nucleation becomes observable at a lower magnetic field of 0.25 mT (Figure 2 (c)). Furthermore, an increase in nucleation site density is evidenced for a positive voltage (+3.0V), while a reduction in the nucleation site density is observed for a negative voltage (-2.5V) (Figure 2 (b)). These results are consistent with a modulation of PMA under voltage but no transition to in-plane magnetization is observed at the highest applied voltage.

Figure 2 (d) shows $H_c$ variation as a function of the applied voltages $V_G$ when the hysteresis loops are measured during the application of the voltage (black squares), right after the voltage is turned off (red circles) and 180s after the voltage is turned off (blue triangles). The observed change in $H_c$ for the measurement performed during the continuous application of the voltage shows a linear response with $V_G$, with $H_c$=-0.15*$V_G$+0.82. For $V_G$=-2.5V the coercivity is enhanced by 53% while for $V_G$=+3.0V it is reduced by 52%. When $H_c$ is measured after $V_G$ is turned off, no significant change with respect to the zero-field state is observed indicating the presence of a volatile effect.

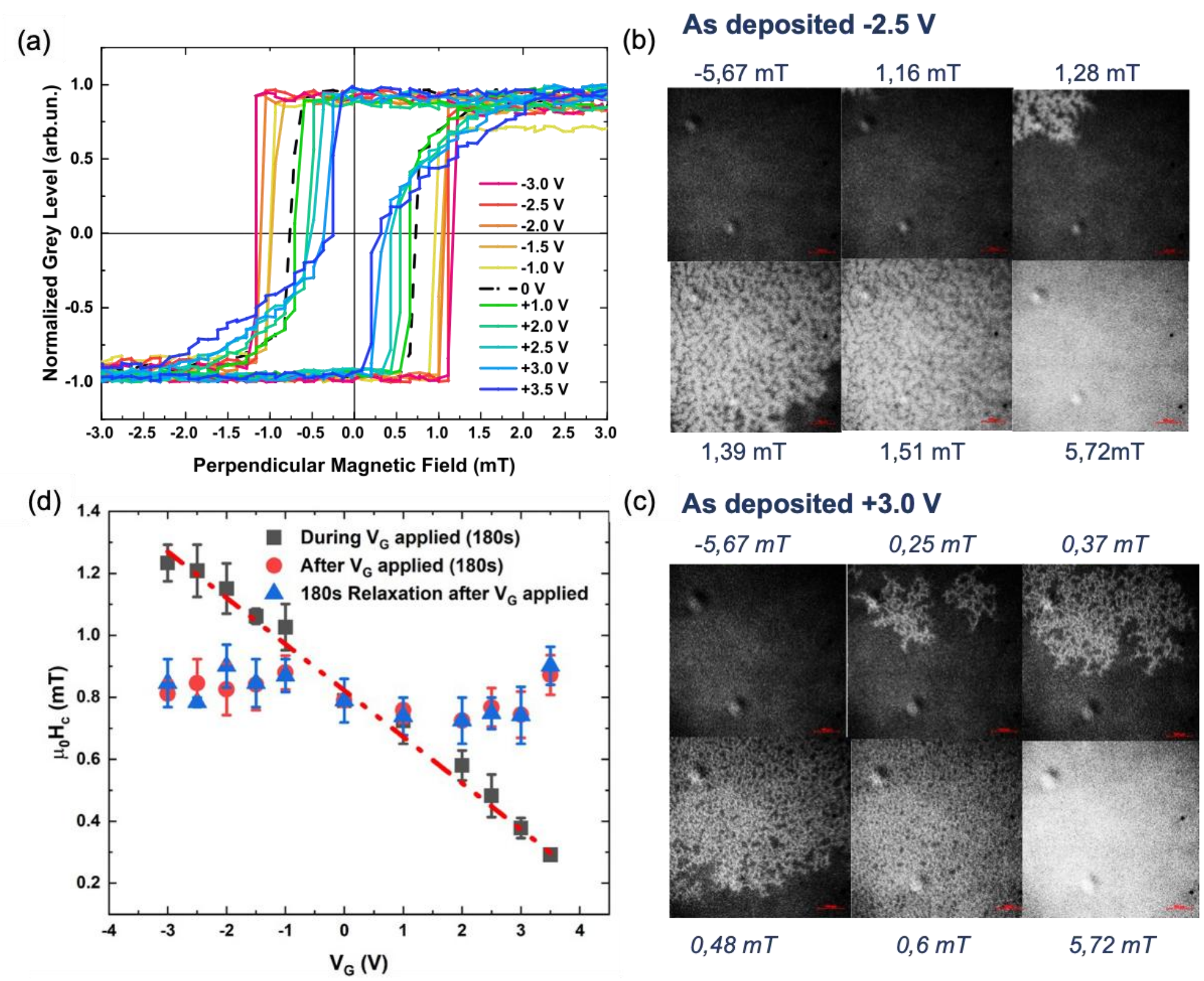

**Figure 2.** Charge- mediated magnetoelectric effect in as-deposited W-CoFeB-MgO samples (a) hysteresis loops as a function of voltage measured during the application of the voltage (duration of 180s), (b)-(c) Domain wall images (pMOKE) for as deposited samples at -2.5 V and 3V at different magnetic fields. (d) Coercivity as a function of the voltage extracted from the hysteresis loops measured during the application of the voltage (black), right after the voltage is turned off (red) and 180s after the voltage is turned off (blue).

These results indicate that the perpendicular magnetic anisotropy of the as-deposited W-CoFeB-MgO systems can be changed linearly under $V_{G,}$ with a purely volatile effect. These results are consistent with the well-known magnetoelectric effect due to the accumulation or depletion of screening electronic charges at the interface between the MgO dielectric layer and the magnetic material. E-field induced charge accumulation/depletion modifies the relative occupancy of Co/Fe-O 3d-2p orbitals at the CoFeB-MgO interface, which are responsible for the interfacial anisotropy energy modification[30].

***MI effect in annealed W/$Co_{20}Fe_{60}B_{20}$/MgO/$HfO_2$ materials***

Post-growth annealing-induced crystallization of W-CoFeB-MgO is a pre-requisite for achieving high PMA and high TMR in CoFeB free layer-based MTJs. The annealing at temperatures higher than 300°C induces Boron (B) migration toward the Tungsten (W) layer which acts as a sink inducing the crystallization of the CoFe layer from the MgO (001) interface, enhancing the PMA at the W/CoFeB[31–33] and CoFeB/MgO interface[34,35]. Figure 3(a) show the hysteresis loops measured by pMOKE for the films annealed at 370°. The square hysteresis loop obtained for an annealing temperature of 370°C is consistent with the presence of PMA most likely induced by the crystallization of the CoFeB and MgO layer[36,37].

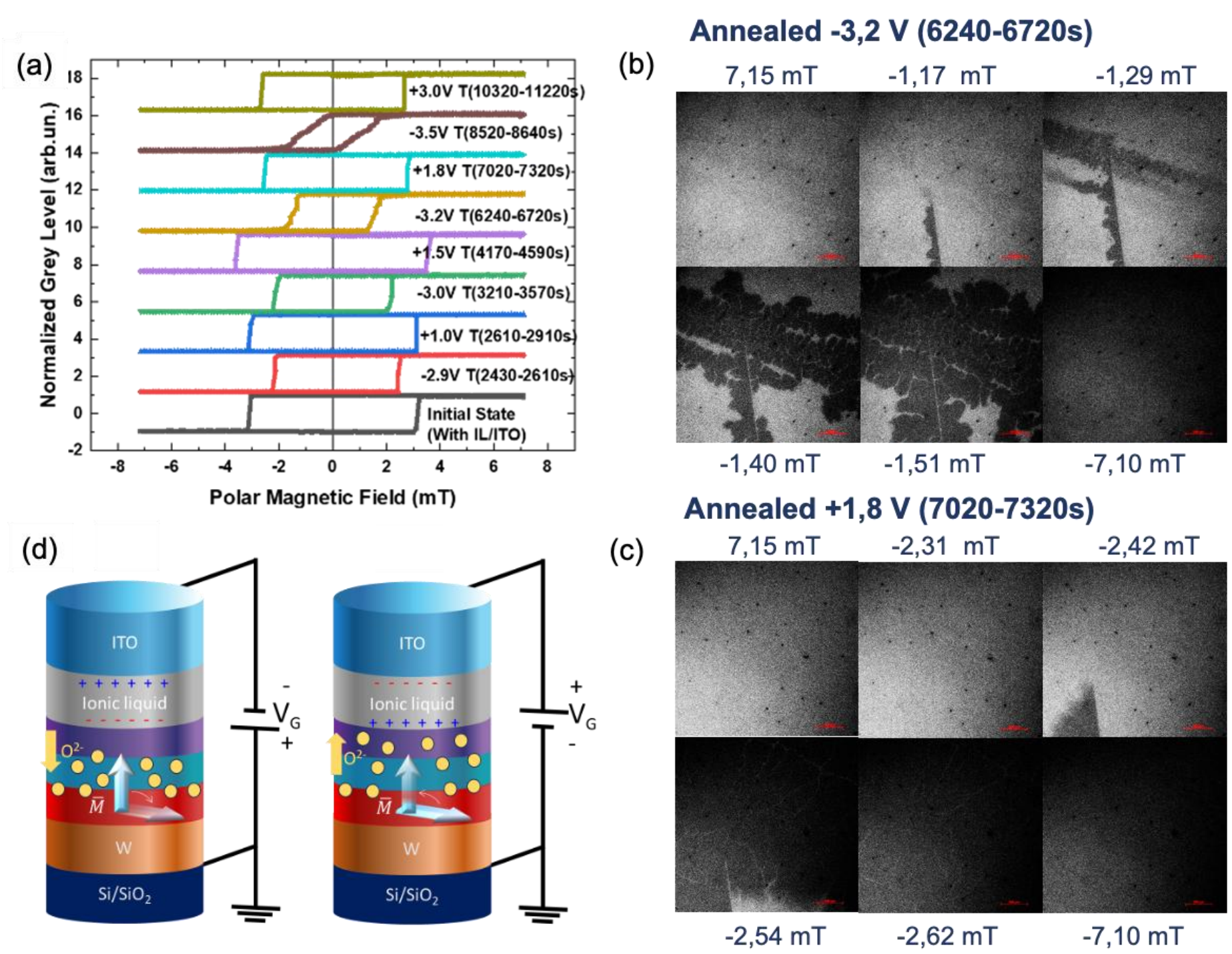

**Figure 3.** Magneto-Ionics effect in W/$Co_{20}Fe_{60}B_{20}$/MgO/$HfO_2$ ultra-thin film annealed at 370°C. (a) Polar hysteresis loops obtained by pMOKE after applying voltage gating for a certain time. Polar MOKE images recorded during a hysteresis measurement after application of (b) negative (-3.2V) and (c) positive (+1.8V) voltages. (d) A schematic showing a possible ion migration scenario under negative and positive voltages.

In Figure 3 (a), hysteresis loops are shown after the application of positive and negative electric fields for the sample annealed at 370°C. A clear modulation of the PMA is observed with an increase (decrease) of $H_c$ for positive (negative) voltages. Application of large negative voltages, up to -3.5V, induces a substantial reduction in perpendicular magnetic anisotropy (PMA), leading to a loss of squareness in the hysteresis loop. In contrast to the results observed in amorphous samples, the E-field effect here is non-volatile and reversible. It is important to note that the polarity of the voltage exhibits an opposite effect compared to the charge-mediated magnetoelectric effect, which depends on the spin-dependent charge screening. These results are consistent with a magneto-ionic effect due to oxygen ions moving reversibly back and forth from/to the CoFeB-MgO interface (see Figure 3 (d)).

MOKE measurements conducted under magnetic fields indicate that domain wall nucleation and propagation are favored at lower fields for negative voltages, as illustrated in Figure 3 (b). Conversely, domain wall nucleation and propagation become less favorable after the application of a positive voltage, as demonstrated in Figure 3 (c). Remarkably, the dendritic domain wall propagation observed after the application of negative voltage (Figure 3 (b)) disappears by the application of a positive voltage (Figure 3 (c)), indicating an increase of interfacial anisotropy and hence a reversible defects control.

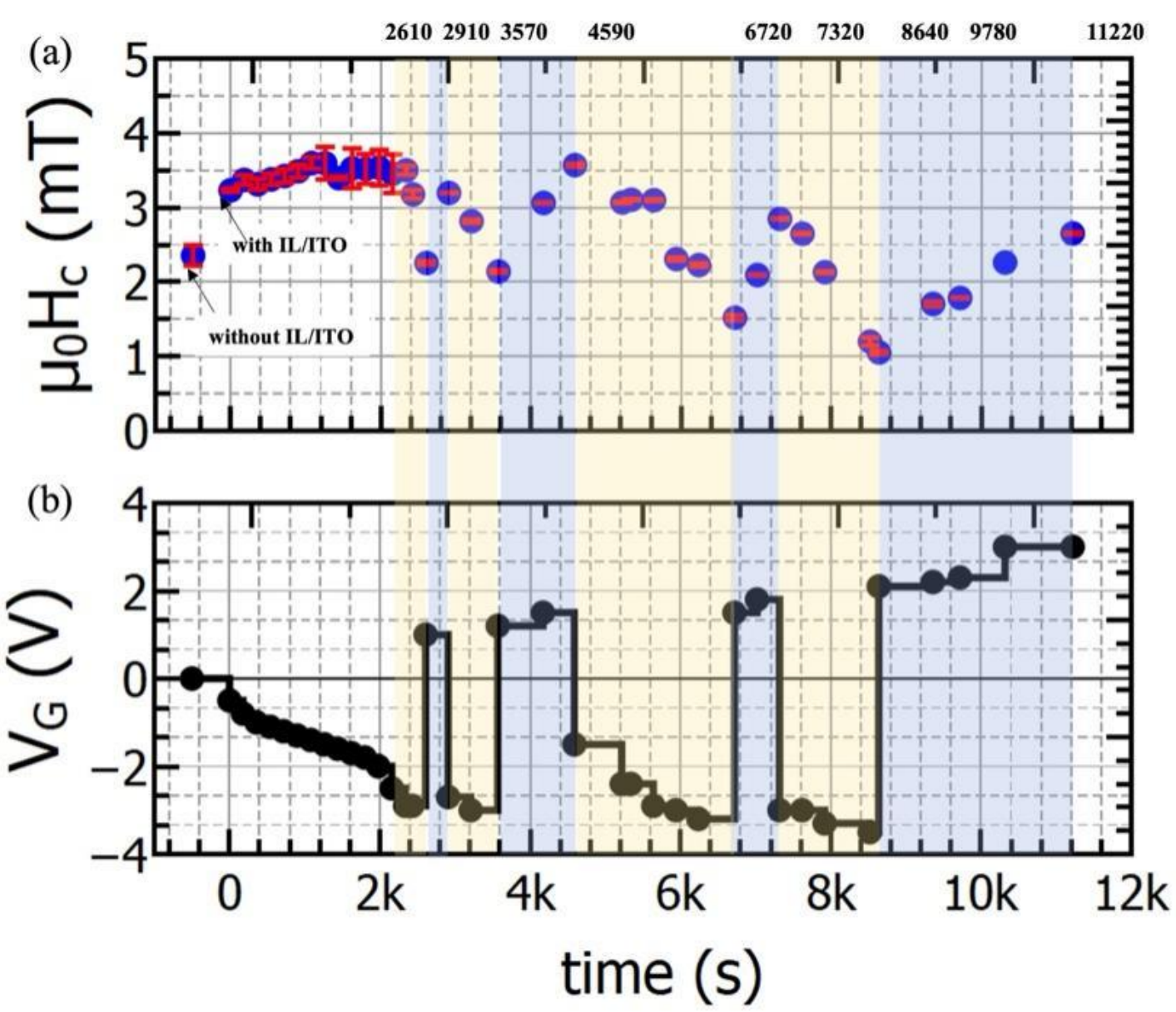

**Figure 4.** Magneto-ionic effect in $W/Co_{20}Fe_{60}B_{20}/MgO/HfO_2$ ultra-thin film annealed at 370°C. The corresponding coercivities and the applied gate voltage versus time of application. Yellow (blue) regions correspond to a negative (positive) $V_G$.

In Figure 4, by alternately applying positive and negative voltages, we can observe a reversible non-volatile modulation of coercivity. The maximum coercivity ($H_c$) achieved is 3.22 mT and the minimum 1 mT, which represents a substantial modulation of magnetic properties. Note that when the samples have been cycled several times over a few days, the amplitude of modulation strongly decreases and only a remaining magneto-electric effect remains. This might be induced by that stable overoxidized ferromagnet is inevitably formed during each gating cycle.

## Discussion

The key finding here is a reversible and non-volatile Magneto-ionic effect in crystallized CoFeB-MgO materials compared to the amorphous state, where only a charge-mediated magneto-electric effect is observed. In comparison with previous studies on annealed structures containing a W buffer layer W/$Co_{20}Fe_{60}B_{20}$/$HfO_2$,[12,28] which exhibited partial reversibility, the stack W/$Co_{20}Fe_{60}B_{20}$/MgO/$HfO_2$ demonstrates full reversibility. In both cases, the $HfO_2$ layer serves as an $O^{2-}$ reservoir. A possible explanation for the difference between annealed CoFeB-MgO-$HfO_2$ and CoFeB-$HfO_2$ structures is that, in annealed W/$Co_{20}Fe_{60}B_{20}$/$HfO_2$ materials, due to the amorphous character of the $HfO_2$ layer, only partial crystallization of the CoFeB layer can be obtained, contrary to the case of the CoFeB-MgO interface, where MgO (001) acts as a template for full crystallization at high temperatures. These distinct features highlight the importance of the crystalline state as a key parameter in promoting ion migration. This is likely attributed to the presence of grain boundaries (typically 10nm) in annealed CoFeB-MgO layers, which can act as reversible ion conduction channels. Another important feature is that the CoFeB/MgO interfacial state plays a significant role in the magneto-ionic response and reversibility. According to first principle calculations[37,42], PMA at the FM/oxide interface is attributed to the FM 3d-O 2p orbital hybridization and the maximum PMA value can be obtained when the interface is fully oxidized, whereas over-oxidization and under-oxidization at the interface reduce the PMA value. The as-deposited interface stays under-oxidized, while the interfacial state can be enhanced after annealing[40,41]. This is consistent with the results that the coercive field after annealing shown in Figure 4 is larger than that of the as-deposited state in Figure 2. Compared with the defects-rich interfacial state in the as-deposited system, the enhanced interfacial state after annealing potentially enables the voltage driven oxygen ions to oxidize the CoFeB layer further. Correspondingly, the results shown in Figure 4 illustrate that the magneto-ionics is not activated until reaching a negative voltage of -2.9 V, while no modification of coercivity occurs at higher voltages (-3.0V) as shown in Figure 2 (d), indicating a lower diffusion barrier of the oxygen ions in the annealed system than the as-deposited system. The reversibility is reached most likely in the over-oxidation state where O ions could potentially be moved back and forth reversibly along the grain boundaries.

From a thermodynamic standpoint, the underlying reason for the full reversibility of the Magneto-Ionic(MI) effect in the annealed W/CoFeB/MgO/$HfO_2$, in contrast to annealed W/CoFeB/$HfO_2$ system[12], can be related to the enthalpy of formation of MgO and $HfO_2$. MgO exhibits a comparatively high enthalpy of formation (-601.8 kJ/mol[42]) compared to FeO (-272.04 kJ/mol[42]) and $Fe_2O_3$ (-825.5 kJ/mol[42], while, the enthalpy of formation of $HfO_2$ (-267.4 kJ/mol[42]) is relatively low. A higher enthalpy of formation typically signifies greater stability and oxygen affinity. Therefore, MgO can serve as a more effective oxygen reservoir when driving oxygen ions from $FeO_x$ by applying a positive voltage, as compared to $HfO_2$.

**Conclusion**

In conclusion, we have observed a charge-mediated magneto-electric effect in amorphous $W/Co_{20}Fe_{60}B_{20}/MgO/HfO_2$ materials induced by the accumulation or depletion of electrons at the CoFeB/MgO interface. After post-deposition annealing at 370°C that induces crystallization of the CoFeB material, we have observed a non-volatile and reversible MI effect. We explain this result by the presence of grain boundaries and the enhanced CoFeB/MgO interfacial state in the annealed sample which favors ionic migration, and by the MgO layer with a relatively high oxygen affinity which favors the reversibility behavior at the CoFeB-MgO interface. The possibility to control ion migration in crystallized CoFeB-MgO materials opens the route to integrate the magneto-ionics effect in CoFeB-MgO based magnetic tunnel junctions.

## Acknowledgements

This project has received funding from the European Union's Horizon 2020 research and innovation program under the Marie Skiodowska-Curie grant agreement No. 861145.